\documentclass{llncs}

\usepackage[utf8]{inputenc}
\usepackage[T1]{fontenc}
\usepackage{underscore}
\usepackage{xcolor}
\usepackage{listings}
\newcommand{\term}[1]{\textbf{#1}}
\newcommand{\libname}[1]{\textrm{#1}}
\newcommand{\systemname}[1]{\textsf{#1}}
\newcommand{\mizar}[0]{\systemname{Mizar}}
\newcommand{\automath}[0]{\systemname{Automath}}
\newcommand{\coq}[0]{\systemname{Coq}}
\newcommand{\hollight}[0]{\systemname{HOL light}}
\newcommand{\camlpfour}[0]{\systemname{Camlp4}}

\newcommand{\toolname}[1]{\textsl{#1}}

\newcommand{\lisppars}[0]{\toolname{lisppars}}
\newcommand{\pvs}[0]{\systemname{PVS}}
\newcommand{\imps}[0]{\systemname{IMPS}}
\newcommand{\acltwo}[0]{\systemname{ACL2}}

\newcommand{\bison}[0]{\toolname{bison}}
\newcommand{\gcc}[0]{\toolname{gcc}}
\newcommand{\pdflatex}[0]{\toolname{pdflatex}}

\usepackage{hyperref}
\hypersetup{
  pdftitle={New developments in parsing Mizar},
  pdfauthor={Czeslaw Bylinski and Josef Urban},
  pdfkeywords={interactive theorem proving, parsing, normalization, text transformation, semantics, syntax},
  colorlinks=true,
  linkcolor=black,
  citecolor=black,
}

\begin{document}

\title{New developments in parsing \mizar{}}
\titlerunning{Parsing \mizar}
\author{Czes\l{}aw Bylinski\inst{1}
\and
Jesse Alama\inst{2}
\thanks{Supported by the ESF
research project \emph{Dialogical Foundations of Semantics} within
the ESF Eurocores program \emph{LogICCC} (funded by the Portuguese
Science Foundation, FCT LogICCC/0001/2007).  Research for this paper was partially done while a visiting
fellow at the Isaac Newton Institute for the Mathematical Sciences in
the program `Semantics \& Syntax'.  Karol P{\k{a}}k deserves thanks for his patient assistance in
   developing customized \mizar{} text rewriting tools.}}
\institute{Department of Programming and Formal Methods\\University of Bia\l{}ystok\\Poland\\\email{czeslaw@mizar.org}}
\institute{Center for Artificial Intelligence\\New University of Lisbon\\Portugal\\\email{j.alama@fct.unl.pt}}

\authorrunning{Bylinski and Alama}

\maketitle

\begin{abstract}
  The \mizar{} language aims to capture mathematical vernacular by
  providing a rich language for mathematics.  From the perspective of
  a user, the richness of the language is welcome because it makes
  writing texts more ``natural''.  But for the developer, the richness
  leads to syntactic complexity, such as dealing with overloading.

  Recently the \mizar{} team has been making a fresh approach to the
  problem of parsing the \mizar{} language.  One aim is to make the
  language accessible to users and other developers.  In this paper we
  describe these new parsing efforts and some applications thereof,
  such as large-scale text refactorings, pretty-printing, HTTP parsing
  services, and normalizations of \mizar{} texts.
\end{abstract}

\section{Introduction}\label{sec:introduction}

The \mizar{} system provides a language for declaratively expressing
mathematical content and writing mathematical proofs.  One of the
principal aims of the \mizar{} project is to capture ``the
mathematical vernacular'' by permitting authors to use linguistic
constructions that mimic ordinary informal mathematical writing.  The
richness is welcome for authors of \mizar{} texts.  However, a rich,
flexible, expressive language is good for authors can lead to
difficulties for developers and enthusiasts.  Certain experiments with
the \mizar{} language and its vast library of formalized mathematical
knowledge (the \mizar{} Mathematical Library, or MML), naturally lead
to rewriting \mizar{} texts in various ways.  For some purposes one
can work entirely on the semantic level of \mizar{} texts; one may not
need to know precisely what the source text is, but only its semantic
form.  For such purposes, an XML presentation of \mizar{} texts has
long been available~\cite{urban2005xmlizing}.  However, for some tasks
the purely semantic form of a \mizar{} text is not what is wanted.
Until recently there has been no standalone tool, distributed with
\mizar{}, that would simply parse \mizar{} texts and present the parse
trees in a workable form.\footnote{One parser tool, \lisppars{}, is
  distributed with \mizar{}.  \lisppars{} is mainly used to facilitate
  authoring \mizar{} texts with Emacs~\cite{urban2006mizarmode}; it
  carries out fast lexical analysis only and does not output parse
  trees.}
% Compounding the problem of the inherent difficulty of parsing \mizar{}
% texts and the lack of a standalone parser, \mizar{} itself is, to a
% large extent, a closed-source system.\footnote{The contents of the
%   \mizar{} Mathematical Library, on the other hand, are now governed
%   by an open-source license~\cite{alama2011licensing}.}  One could try
% to reverse-engineer \mizar{} to carry out private experiments, but
% this is a steep obstacle.  One can request assistance from the
% \mizar{} developers for specific tasks, but this is
% obviously an unacceptable method for solving one's own problems.  And
% even if one has access to the \mizar{} codebase, it may not be clear
% how to carry out a specific rewriting task.

Parsing texts for many proof assistants is often facilitated through
the environment in which these proof assistants are executed.  Thus,
texts written for those systems working on top of a Lisp, such as
\imps, \pvs, and \acltwo, already come parsed, so one has more or less
immediate access to the desired parse trees for terms, formulas,
proofs, etc.  Other systems, such as \coq{} and \hollight{}, use
syntax extensions (e.g., \camlpfour{} for Objective Caml) to ``raise''
the ambient programming language to the desired level of proof texts.
For \mizar{}, there is no such ambient environment or read-eval-print
loop; working with \mizar{} is more akin to writing a C program or
\LaTeX{} document, submitting it to \gcc{} or \pdflatex{}, and
inspecting the results.

% It must be admitted, then, that in comparison to comparable
% proof assistants, \mizar{} does pose obstacles for the potential
% enthusiast, student, teacher, or developer who wishes to work with
% \mizar{} texts in ways that go beyond the customary write-verify-edit
% loop.

% Toward a release of their (presently closed-source) code under a
% standard open-source license, the \mizar{} team is reorganizing parts
% of the the \mizar{} toolchain, starting with the parser.
This paper describes new efforts by the \mizar{} team to make their language
more workable and illustrates some of the fruits these efforts have
already borne.  This paper does not explain \emph{how} to parse
arbitrary \mizar{} texts.  And for lack of space we cannot go into the
detail about the \mizar{} system;
see~\cite{grabowski2010mizar,matuszewski2005mizar}.%   The scope of this
% paper is to announce new developments by the \mizar{} team that make
% parsing of \mizar{} texts more accessible for users and developers.

% % In Section~\ref{sec:challenges} we briefly relay some of the
% % challenges of parsing \mizar{} texts.
In Section~\ref{sec:layers}, we discuss different views of \mizar{}
texts that are now available.  Section~\ref{sec:applications}
describes some current applications made possible by opening up
\mizar{} texts, and describes some HTTP-based services for those who
wish to connect their own tools to \mizar{} services.
Section~\ref{sec:conclusion} concludes by sketching further work and
potential applications.

%\section{Challenges}\label{sec:challenges}

\section{Layers of a \mizar{} text}\label{sec:layers}

% The richness of the \mizar{} language allows one to define various
% levels at which one can view a \mizar{} text.  That there are
% different views is, of course, obvious (conisder, for example, the
% classical distinction between tokenizing a stream of text and
% analyzing a sequence of tokens).  In the case of \mizar{}, though, one
% can go specify further possibilities, permitting further increasing
% levels of ``meshing'' of syntax and (a kind of) semantics.

It is common in parsing theory to distinguish various analyses or
layers of a text, considered in the first place as a sequence of bytes
or characters~\cite{aho2007compilers}.  Traditionally the first task
in parsing is \term{lexical analysis} or \term{scanning}: to compute,
from a stream of characters, a stream of \emph{tokens}, i.e.,
terminals of a production grammar $G$.  From a stream of tokens one
then carries out a \term{syntactic analysis}, which is the synthesis
of tokens into groups that match the production rules of $G$.

% In the case of the \mizar{} language, there are difficulties at both
% levels of parsing.

% A problem common to both levels is that
One cannot, in general, lexically analyze \mizar{} texts without
access to the MML.  Overloading (using the same symbol for
multiple, possibly unrelated meanings) already implies that parsing
will be non-trivial, and overloading is used extensively in the
\mizar{} library.  Even with a lexical analysis of a \mizar{} text,
how should it be understood syntactically?  Through \mizar{}'s support
for \term{dependent types}, the overloading problem is further
complicated.  Consider, for example, the \mizar{} fragment
\begin{lstlisting}
let X be set,
    R be Relation of X, Y;
\end{lstlisting}
The notion of a (binary) relation is indicated by the non-dependent
(zero-argument) type \verb+Relation+.  There is also the binary notion
\emph{relation whose domain is a subset of $X$ and whose range is a
  subset of $Y$}, which is expressed as \verb+Relation of X,Y+. Finally, we have the one-argument notion \emph{relation whose
  domain is a subset of $X$ and whose range is a subset of $X$} which
is written \verb+Relation of X+. In the text fragment above, we have
to determine which possibility is correct, but this information would
not contained in a token stream (is \verb+Y+ the second argument of an
instance of the binary \verb+Relation+ type, or is it the third
variable introduced by the \verb+let+?).

\subsection{Normalizations of \mizar{} texts}\label{sec:strictness}

One goal of opening up the \mizar{} parser is to help those interested
in working with \mizar{} texts to not have to rely on the \mizar{}
codebase to do their own experiments with \mizar{} texts.  % The
% \mizar{} parser has up to now been part of the \mizar{} codebase in
% the form of library routines rather than as a standalone executable;
% no tool of the standard \mizar{} distribution has provided just an
% interface to the parser.
We now describe two normalizations of (arbitrary) \mizar{} texts,
which we call weakly strict and more strict.  The results of these two
normalizations on a \mizar{} text can be easily parsed by a standard
LR parser, such as those generated by the standard tool
\bison\footnote{\url{http://www.gnu.org/software/bison/}} and have
further desirable syntactic and semantic properties.  Other
normalizations beyond these two are certainly possible.  For example,
whitespace, labels for definitions, theorems, lemmas, etc., are
rewritten by the normalizations we discuss; one can imagine
applications where such information ought not be tampered with.

\subsection{Weakly strict \mizar}
\label{sec:weakly-strict}

The aim of the weakly strict \mizar{} (WSM) transformation is to
define a class of \mizar{} texts for which one could easily write an
standard, standalone parser that does not require any further use of
the \mizar{} tools.  In a weakly strict \mizar{} text all notations
are disambiguated and fully parenthesized, and all statements take up
exactly one line.  (This is a different transformation than
single-line variant \libname{AUT-SL} of the \automath{}
system~\cite{debruijn1971autsl}.)  Consider:
\begin{lstlisting}
reserve P,R for Relation of X,Y;
\end{lstlisting}
This \mizar{} fragment is ambiguous: it is possible that the variable
\verb+Y+ is a third reserved variable (after the variables \verb+P+
and \verb+R+), and it is possible that \verb+Y+ is an argument of the dependent type \verb+Relation of X,Y+.  The text becomes disambiguated by the weakly
strict \mizar{} normalization to
\begin{lstlisting}
reserve P , R for ( Relation of X , Y ) ;
\end{lstlisting}
and now the intended reading is syntactically evident, thanks to
explicit bracketing and whitespace.  (Any information that is
implicitly contained by whitespace structure in the original text is
destroyed.)

The result of the one-line approach of the weakly strict \mizar{}
normalization is, in many cases, excessive parenthesization,
unnecessary whitespace, and rather long lines.\footnote{The longest
  line in the ``WSM-ified'' library has length $6042$.  About $60$\%
  (to be precise, $694$) of the articles in the WSM form of the
  current version of the \mizar{} Mathematical Library ($4.181.1147$)
  have lines of length at least $500$ characters.  The average line
  length across the whole ``WSM-ified'' library is $54.7$.} The point
of the weakly strict \mizar{} normalization is not to produce
attractive human-readable texts.  Instead, the aim is to transform
\mizar{} texts so that they have a simpler grammatical structure.

\subsection{More Strict \mizar}
\label{sec:more-strict}

A second normalization that we have implemented is called, for lack of
a better term, more strict \mizar{} (MSM).  The aim of the MSM
normalization is to to define a class of \mizar{} texts that are
canonicalized in the following ways:
\begin{itemize}
\item From the name alone of an occurrence of a variable one can
  determine the category (reserved variable, free variable, bound
  variable, etc.) to which the occurrence belongs.  (Such inferences
  are of course not valid for arbitrary \mizar{} texts.)
\item All formulas are labeled, even those that were unlabeled in the
  original text.
\item Some ``syntactic sugar'' is expanded.
\item Toplevel logical linking is replaced by explicit
  reference. Thus,
  \begin{lstlisting}
@$\phi$@; then @$\psi$@;
  \end{lstlisting}
  using the keyword \verb+then+ includes the previous statement
  ($\phi$) as the justification of $\psi$. Under the MSM
  transformation, such logical relationships are rewritten as
  \begin{lstlisting}
Label1: @$\phi$@;
Label2: @$\psi$@ by Label1;
  \end{lstlisting}
  Now both formulas have new labels \verb+Label1+ and \verb+Label2+.
  The logical link between $\phi$ and $\psi$, previously indicated by
  the keyword \verb+then+, is replaced by an explicit reference to the
  new label (\verb+Label1+) for $\phi$.
\item All labels of formulas and names of variables in a \mizar{} are
  serially ordered.
\end{itemize}
% Computations corresponding to these normalizations are carried out
% internally by the \mizar{} verifier.  Internally, for example, a
% serial ordering of variables and expands all syntatic sugar.  The MSM
% transformation spells out some of these ``semantic'' computations.
% and outputs not a proof object, but another \mizar{} text the
% expresses these relationships.

% The advantage of the MSM \mizar{} transformation over a plain \mizar{}
% text, and a WSM text, is that it frees the \mizar{} developer from
% having to write complex, error-prone procedures to infer which of the
% various roles that could be played by an occurrence of a variable it
% is actually playing.
MSM \mizar{} texts are useful because they permit certain ``semantic''
inferences to be made simply by looking at the syntax.  For example,
since all formulas are labeled and any use of a formula must be done
through its label, one can infer simply by looking at labels of
formulas in a text whether a formula is used.  By looking only at the
name of a variable, one can determine whether it was introduced inside the current proof or was defined
earlier.  % One can infer simply from the name of two variables

\section{Applications}
\label{sec:applications}

Opening up the \mizar{} parser by providing new tools that produce
parse trees naturally suggests further useful text transformations,
such as pretty printing.  An HTTP parsing service for these new
developments is available for public consumption.  Four services are
available.  Submitting a suitable \verb+GET+ request to the service
and supplying a \mizar{} text in the message body, one can obtain as a
response the XML parse tree for the text, a pretty-printed form of it,
or the WSM or MSM form of a text (either as plain text or as XML).
The HTTP services permit users to parse \mizar{} texts without having
access to the MML, or even the \mizar{} tools.  See
\begin{quote}
  \url{http://mizar.cs.ualberta.ca/parsing/}
\end{quote}
to learn more about the parsing service, how to prepare suitable HTTP
parsing requests, and how to interpret the results.

% (The HTTP services are also stopgap measure until \wsmparser{}
% and \msmprocessor{} belong to the standard \mizar{} distribution.)

% Unlike other web services for working with \mizar{} texts, such as
% \MizAR{}~\cite{urban2010automated} and the \mizar{}
% wiki~\cite{urban2010wiki}, our parsing is essentially only a do not deal with (complete) semantic
% representations of \mizar{} texts.  % The service reports only the the
% % most flagrant kinds of syntactic errors that could afflict a \mizar{}
% % text, such as having a malformed environment, containing non-ASCII
% % characters, or illegally combining keywords.
% A \verb+200 OK+ response
% from the service by no means ensures that the given \mizar{} text is
% logically or mathematically meaningful or valid.

\section{Conclusion and Future Work}\label{sec:conclusion}

Parsing is an essential task for any proof assistant.  In the case of
\mizar{}, parsing is a thorny issue because of the richness of its
language and its accompanying library. New tools for parsing \mizar{},
with an eye toward those who wish to design their own \mizar{}
applications without (entirely) relying on the \mizar{} tools, are now
available.  Various normalizations for \mizar{} texts have been
defined.  Further useful normalizations are possible.  At present we
are experimenting with a so-called ``without reservations'' \mizar{}
(WRM), in which there are no so-called reserved variables; in~WRM
texts the semantics of any formula is completely determined by the
block in which it appears, which should make processing of \mizar{}
texts even more efficient.

\bibliographystyle{splncs03}
\bibliography{parsing-mizar}

\end{document}